\documentclass[letterpaper]{article}
\usepackage{aaai}
\usepackage{times}
\usepackage{helvet}
\usepackage{courier}
\usepackage{amsmath}
\usepackage{booktabs} 
\usepackage{multirow}
\usepackage{aaai}
\usepackage{graphicx} 
\usepackage{indentfirst}
\usepackage{url}
\begin{document}
%
\title{Measuring Female Representation and Impact in Films over Time}

\author{\Large \textbf{Luoying Yang \textsuperscript{\rm 1}, Zhou Xu \textsuperscript{\rm 2}, Jiebo Luo\textsuperscript{\rm 3}}\\ 
\textsuperscript{\rm}University of Rochester\\ 
luoying\_yang@urmc.rochester.edu \textsuperscript{\rm 1}, zxu17@ur.rochester.edu \textsuperscript{\rm 2}, jluo@cs.rochester.edu \textsuperscript{\rm 3}\\ 
}

\maketitle
\begin{abstract}
\begin{quote}
Women have always been underrepresented in movies and not until recently has the representation of women in movies improved. To investigate the improvement of female representation and its relationship with a movie's success, we propose a new measure, the female cast ratio, and compare it to the commonly used Bechdel test result. We employ generalized linear regression with $L_1$ penalty and a Random Forest model to identify the predictors that influence female representation, and evaluate the relationship between female representation and a movie's success in three aspects: revenue/budget ratio, rating, and popularity. Three important findings in our study have highlighted the difficulties women in the film industry face both upstream and downstream. First, female filmmakers, especially female screenplay writers, are instrumental for movies to have better female representation, but the percentage of female filmmakers has been very low. Second, movies that have the potential to tell insightful stories about women are often provided with lower budgets, and this usually causes the films to in turn receive more criticism. Finally, the demand for better female representation from moviegoers has also not been strong enough to compel the film industry to change, as movies that have poor female representation can still be very popular and successful in the box office.
\end{quote}
\end{abstract}

\section{Introduction}
Film is a common entertainment form that fulfills the audience's desire to make emotional connections with characters and learn about their social world. However, it was very rare for women to see inspiring counterparts on the big screen in the past. Many key roles in film-making, such as directors and cinematographers, were for many decades dominated almost entirely by men ~\cite{lauzen2012celluloid}, and women did not have enough power to make demands in the film industry. Consequently, women have been constantly underrepresented in movies. Even when they are present, women are often portrayed in circumscribed and subordinated ways in traditionally feminine (i.e., stereotyped) roles, such as nonprofessionals, homemakers, wives or parents, and sexual gatekeepers ~\cite{Collins2011}. Lacking a role model on the big screen is detrimental for young girls. They are discouraged from pursuing ambitions and participating actively in social affairs ~\cite{womenpolitical}. Therefore, female under-representation is a critical issue that must be addressed. 

In more recent times, women have made inroads into various fields and films have started to respond to female viewers with strong and well-rounded female characters \cite{murphy2015role,heldman2016}. There are many studies and projects devoted to studying evolving feminism in films, centering on both the \textit{upstream effects}, in which content is structured through the actions of major filmmakers in gendered organizations who presume the public's preferences, as well as the \textit{downstream effects}, where audiences respond to content and attitudes are formed and reinforced. In our work of examining gender representation in films, we ask the following two questions: What factors in the film-making process have a significant impact on female representation in films, and do films that feature more female representation outperform those that do not commercially?

To answer these questions, we need to define a proper measure for female representation in films. The Bechdel test is a popular measure for examining how well-rounded and complete the representations of women in media are. The test asks only three questions: 
\begin{itemize}
    \item Are there at least two women in the film who have names?
    \item Do those women talk to each other?
    \item Do they talk to each other about something other than a man?
\end{itemize}
The test is simple and has been widely used in many film gender studies to measure female representation\cite{lindner2017movies,lindner2015million}. However, one major criticism of the test is that it fails to reveal the hidden gender imbalance structure\cite{micic2015female}. A movie can pass the Bechdel test yet still portray women as auxiliary characters with minimal screen time. For example, both Wonder Woman and The Martian passed the test even though the importance of female characters is very different between the two movies. Therefore, we propose a new measure of female representation in films: the percentage of female cast members in the whole cast. This proposal is based on the intuition that if more women are cast for a film, then women will have more representation on the big screen. 

In addition to a new measure of female representation, we would also like to consider different aspects of movie success. Box office return is a popular measure of movie success because it is directly linked to profit. However, movie ratings and popularity are also important measures of a movie's success. Public acclamation and popularity brings confidence to the studios and producers to inspire similar films or sequels even when the box office return is not substantially high. 

Our analyses are completed in two steps. Drawing on a sample of the most widely distributed films, we first combine a content analysis using the Bechdel Test with film-making data such as the budget and gender of crews to examine what factors are influential on encouraging more female representation during the film-making process. Next, including all film characteristics such as female representation, budget, genre, and many others, we further examine whether better female representation can increase the chance of success for films while adjusting for other possible confounding variables. In doing so, we contribute to sociological theories about the reciprocal impact of feminism.

The remainder of the paper is organized as follows. In section 2, we review the background literature that studies the gender gap in films and analyzes box office performance. In section 3, we describe the data features and the data collection process. In section 4, we present the methods and models used throughout this study. We present our experiments and the corresponding results in Section 4. In section 5, we present and discuss our conclusions and future directions.

\section{Related work}
Most of the studies that center on gender inequality in media are content analysis ~\cite{murphy2015role,heldman2016,micic2015female}, which means they study how women are portrayed in media. While content analysis is beyond the scope of our study, the related works have shown that the roles of women are evolving on the big screen. There are also many studies about female underrepresentation, albeit on a small scale and with simple analysis. Thomas has shown that in the top 10 worldwide highest grossing films of 2016, women speak only 27\% of the lines~\cite{201610}. The finding is novel but the sample size is too small. A team at the Rhodes Information Initiative at Duke University published a report that movies which pass the Bechdel Test have a statistically significant higher return on investment compared to films that do not~\cite{Duke}. The tech company Shift72 published a similar report which states that all films that pass the Bechdel Test surpassed the box office returns of films that fail this test~\footnote{https://shift7.com/media-research/\#arrow-jump}. However, these reports only presented statistics of mere comparisons between movies that pass the test and movies that do not, which are too simple to conclude that “more women in the film means more success (movies are more profitable)." Such analyses failed to adjust for confounding sources such as the genres. Linder et al. have shown that after adjusting for confounding factors such as production budgets, movies with more female representation (passing the Bechdel test) tend to have smaller production budgets and consequently earn less money. However, given the same budget, female representation does not boost the box office return ~\cite{lindner2015million}. They have conducted a similar analysis for movie critics using the same set of movies and obtained a similar conclusion that female representation has little effect on critics \cite{lindner2017movies}. The analysis in these papers is the closest to ours. However, this study only drew samples of 974 films from the 2000-2009 decade and considered only a few confounding variables such as the budget and genres. Our analysis include a larger sample from a longer time span with a larger set of confounding variables to provide deeper insights over time.

\section{Data}
Three data sources are used in this project: the Bechdel Test Movie List \footnote{https://https://bechdeltest.com}, the Internet Movie Database (IMDb) \footnote{https://www.imdb.com/interfaces/}, and the Movie Database (TMDb) \footnote{https://developers.themoviedb.org}. The Bechdel Test Movie List contains over 8000 movies crowd-sourced through the Internet with a flag showing if each of them passes the Bechdel Test or not. IMDb has an up-to-date movie dataset, which is widely used among many movie related projects. However, since IMDb does not provide an official API, we have limited access to some data, such as the full list of the cast and crew members of each movie. TMDb on the other hand is a crowd-sourced online movie database similar to IMDb. It is used less often, but with the advantage of having an API, we can easily access a variety of movie-related data of interest to us.
\subsection{Data Acquisition}
The Bechdel Test Movie List is downloaded using its API. It contains 8190 movies at the time, with 7 fields. The relevant features include imdbid (can be used to join with other datasets), title, year, and rating (0 means no two women, 1 means no talking, 2 means talking about a man, 3 means it passes the test).

The IMDb dataset is downloaded from its website, where only the title.basic table is used. It contains 9 fields, including tconst (imdbid), titleType, primaryTitle, originalTitle, isAdult, startYear, endYear, runtimeMinutes, and genres.

Getting the TMDb dataset requires more work with its API. The first step is to obtain the tmdbids that are associated with the imdbids within the Bechdel Movie List. Then the APIs are called iteratively to access the detail and credits of each movie. By rearranging the tables, we are finally able to get our desired features including: budget, revenue, vote average, vote count, popularity (a comprehensive measure calculated from release data, number of votes, number of views, etc. from TMDb), production companies, number of cast members, and gender of the cast and crew members.

\subsection{Data Preprocessing}
As mentioned from previous sections, we have three data sources: Bechdel Test Movie List, IMDB, and TMDB. The previous two are well established as tabular forms and require minimal processing.

For the TMDb data, new features are created to better represent the movie measurables of interest. First, the revenue-budget ratio is calculated through simple division to represent the profitability of each movie. There are some exceptions: a few non-commercial movies have budgets of 0, so we removed these rare cases from our dataset. Some movies have a long list of production companies, but only those at the top of the list provide major contributions, so we excise the list and set an upper limit of 5 production companies for each movie. In addition, the movie credits features (number, gender, and title of cast and crew members) are aggregated to extract new features such as female cast ratio, number of core crew members (directors, producers and screenplays), and the female core crew ratios.

Finally, the Bechdel Movie List, IMDb, and TMDb datasets are joined according to the imdbid. The categorical variables are then one-hot encoded and the numerical variables are min-max normalized in preparation for modeling. The final table contains 4232 observations and 67 variables.

\section{Methodology}
Our primary goal is to evaluate what factors have an effect on female representation in films and the impact of female representation on films' success. Given this purpose, the primary method of our analysis is an explanatory model which focuses on modeling the true generation of sample data rather than making the best prediction. Many believe that models with high explanatory power are inherently of high predictive power as well. Conflation between explanation and prediction is common, however the type of uncertainty associated with explanation is of a different nature than that associated with prediction \cite{helmer1959epistemology}. Therefore, we add a prediction model as the secondary task to compare the predictive performance with our primary model. For our primary model, we consider Generalized Linear Regression with a penalty term for model selection. Generalized linear regression is one of the most commonly used statistical models for explanatory modeling, and the addition of the penalty term puts constraints on regression coefficient estimation to keep only the relevant variables from the model. It is ideal for our goal of explanatory modeling with variable selection. For our secondary model, we consider Random Forest. Random Forest is known for being robust in predictive modeling with feature ranking by importance. We expect Random Forest to have very satisfactory performance as a predictive model for our data and to use its performance to compare with the predictive performance of our primary model. 
In order to compare the coefficient estimates of different variables from the regression model, variables are normalized by min-max normalization to $[0,1]$ so that they are on the same scale, and the same dataset is fitted to both the regression model and Random Forest model. After obtaining the variable selection and predictors that rank results from the two methods, we compare the set of selected variables and the top ranked predictors to see if they are consistent. For both methods, we split the samples by a 7:3 ratio into the training and test sets and compare the prediction accuracy to see if they have similar performance. 

\subsection{Generalized Linear Regression with $L_1$ Penalty}
Generalized linear regression is a broad class of models that includes regular regression for continuous response as well as models for discrete responses. The relationship between a dependent variable and one or more independent variables can be explained by the magnitude and sign of the regression coefficient estimates. Depending on the types of response variables, we include the regular linear regression for continuous responses and logistic regression for binary responses in our analysis. 
In addition, we also include a $L_1$ regularization term to shrink the coefficient estimates toward 0 so that only highly correlated variables remain in the model \cite{tibshirani1996regression}. A tuning parameter controls the degree of shrinkage such that varying the tuning parameter value results in different estimates. We employ the Bayesian Information Criterion (BIC) to choose the optimal model, which focuses on finding the true model that generates the data. If using an error-based or accuracy-based criterion such as cross-validation for variable selection, when correlated variables enter the model, any one of the correlated variables being selected can give a good predictive performance; however BIC selects the variables that yield the best likelihood for the model and removes the rest, thus serving our purpose of explanatory modeling perfectly. 

\subsection{Random Forest}
The Random Forest is one of the most effective methods for predictive analysis as either a classification algorithm or regression model ~\cite{ho1995random}. By selecting the nodes randomly and aggregating the trees for pooled results, Random Forest is known to be robust to noise and outliers, as well as the value changes of parameters on a fine scale, including the number of trees aggregated and the number of nodes being selected at each node split. Therefore, we do not manually vary the parameter values on a fine scale in order to achieve the optimal results. The model selection, which is the ranking of predictors by their changes in purity, is done internally through computing the Out-of-bag (OOB) error from aggregating the trees. 

\section{Experiments}
In this section, we present the basic features of female representation (i.e. the Bechdel test result and female cast ratio), movie success (i.e. the revenue/budget ratio, movie rating and movie popularity), and our modeling results of their associations. 
\subsection{Preliminary}
\begin{figure*}
    \includegraphics[width=\textwidth]{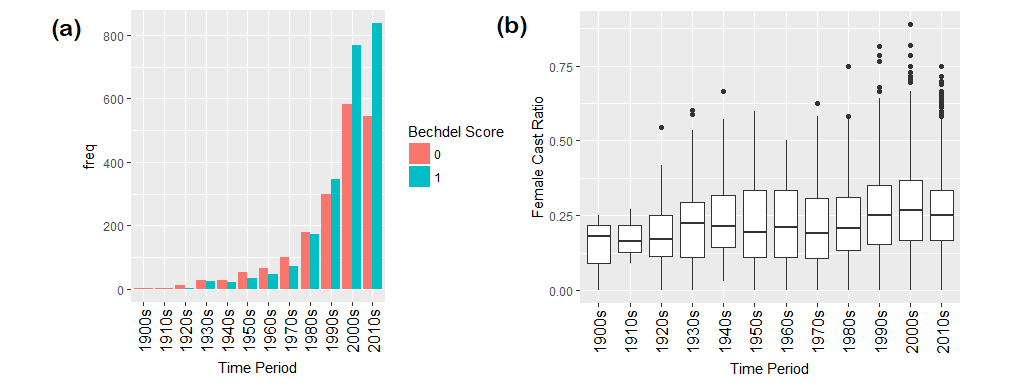}
    \caption{Women representation in movies over time. (a) The distribution of the Bechdel test results over the last 12 decades; (b) The distribution of the female cast ratio over the last 12 decades. }
    \label{Fig 1}
\end{figure*}

We examine the two most important measures of female representation in movies, the Bechdel test result and the female cast ratio. About 55.01\% of the movies in our dataset pass the Bechdel test. As Fig.\ref{Fig 1} shows, in the 21st century, the movies that pass the test have outnumbered the ones that do not. The average female cast ratio is 0.26, and Fig.\ref{Fig 1} shows that the movies in latter time have higher female cast ratios compared to the former time. Using a simple univariate linear regression, we conclude that as time progresses, the number of movies that pass the Bechdel test increases (coefficient estimate = 1.61, p-value$<0.0001$) and more females are cast in the movies (coefficient estimate = 0.11, p-value$<$0.0001). These two measures are highly correlated (t test, p-value$<$0.0001), as movies that have passed the Bechdel test have a higher female cast ratio (0.31) than those which did not (0.20). Among all the variables, we are particularly interested in budget because we want to see whether movies with better female representation were better supported. The results are disappointing in that higher female cast ratio is significantly associated with lower budget (Pearson correlation coefficient = -0.099, p-value$<$0.0001). Similarly, movies that pass the Bechdel test had a significantly lower budget than movies that did not (t test, p-value = 0.001). 

Exploring their association with movie success, we found that the revenue/budget ratio is not significantly associated with the female cast ratio (Pearson correlation coefficient = 0.016, p-value=0.28). Meanwhile, the revenue/budget ratio does not differ significantly between the movies with different Bechdel test results (t test, p-value=0.14). On the other hand, higher movie rating is significantly correlated with lower female cast ratio (Pearson correlation coefficient = -0.15, p-value$<$0.0001). Movie ratings in the group of movies that pass the Bechdel test are significantly lower than the group of movies that do not (t test, p-value$<$0.0001); Similarly, higher movie popularity is also significantly correlated with lower female cast ratio (Pearson correlation coefficient = -0.091, p-value$<$0.0001). Movies that pass the Bechdel test are significantly less popular than the movies that do not. 

\begin{figure}
    \includegraphics[width=0.5\textwidth]{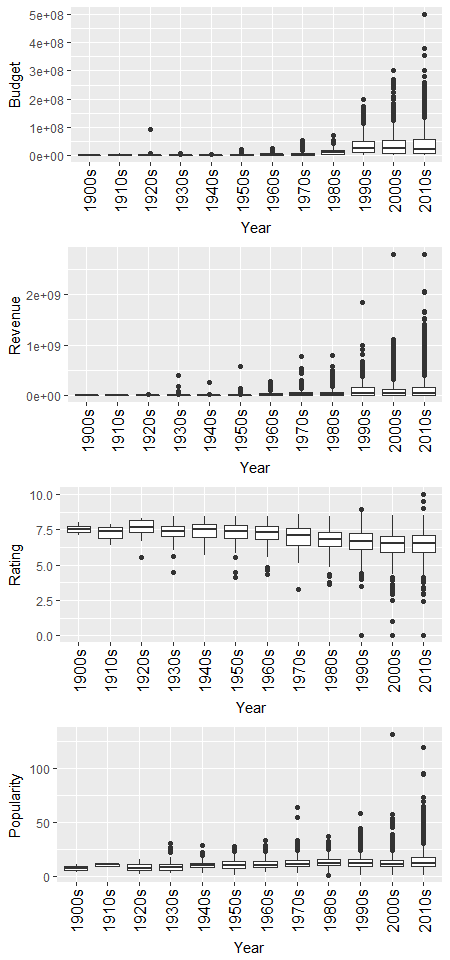}
    \caption{Movie budget, revenue, rating and popularity over the last 12 decades.}
    \label{movie_time}
\end{figure}

The statistical analyses have shown that female representation seems to have a negative impact on movie success. However, we cannot conclude the relationship between movie success and female representation. We have found that female representation is improving over time, and audience reaction to movies is also changing over time. From Fig.\ref{movie_time}, overall, movie rating decreases significantly as time progresses (Pearson correlation coefficient = -0.25, p-value$<$0.0001) and popularity increases significantly as time progresses (Pearson correlation coefficient = 0.097, p-value$<$0.0001). The revenue/budget ratio does not change significantly over time (Pearson correlation coefficient = -0.012, p-value=0.43) because both budget (Pearson correlation coefficient = 0.29, p-value$<$0.0001) and revenue (Pearson correlation coefficient = 0.19, p-value$<$0.0001) increase significantly over time. We further investigate the relationship between movie success and female representation in different time periods. The majority of the movies were produced in the United States, and the feminism movement in the United States reached a peak around 1970. From Fig.\ref{Fig 1}, movies that did not pass the Bechdel test outnumber the movies that did dramatically before 1970, but afterwards there were more movies that passed the test. Before 1970, the relationship between the female cast ratio and movie success is the same as for the overall data. However, all three aspects did not differ significantly between movies that passed the Bechdel test and movies that did not. Going into 1970-2015, the relationship is the same as for the overall data. After 2015, movie rating is no longer significantly correlated with the female cast ratio (Pearson correlation coefficient = -0.078, p-value=0.053), and all three aspects do not differ significantly between movies that passed the Bechdel test and movies that did not. We can see that before 1970 and after 2015, movies that passed the Bechdel test achieved success at a similar level as movies that did not pass the test, while between these two time points movies with better female representation suffered from a lower rating and lower popularity. 

From the preliminary analysis, we can see that there exists significant relationship between female representation and movie success. However, they are also correlated with time as female representation is improving, movie popularity is increasing, and ratings are decreasing. We rely on the regression model and Random Forest regression to adjust for any confounding effects throughout the analyses.

\subsection{The Improvement in Women Representation}
\begin{figure*}
    \includegraphics[width=1.0\textwidth]{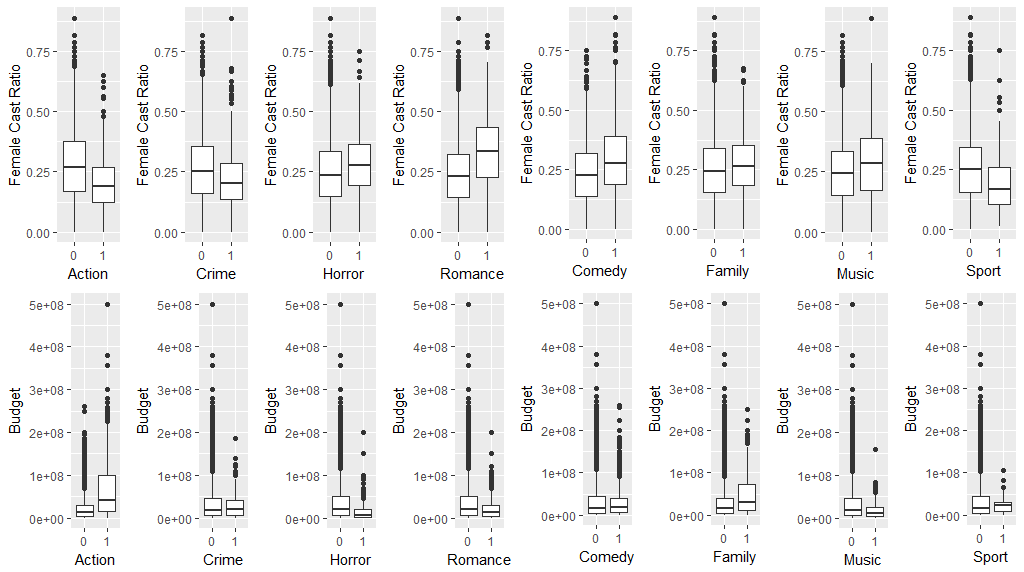}
    \caption{The female cast ratio and budget distribution in the genres selected by the regression model. }
    \label{female_budget}
\end{figure*}

We first evaluate the factors that have strong associations with the Bechdel test result. Setting the Bechdel test result as the response variable of a logistic regression with penalty and a Random Forest, we include the variables of year, adult content, run minutes, budget, number of cast members, female director ratio, female producer ratio, female screenplay writer ratio, 26 production company indicators, and 25 genre indicators as the predictors. The same set of predictors is also fitted to the models with the female cast ratio being the response variable. The variable selection results are shown in Table \ref{Table 1}. For both response variables, the regression model and the Random Forest model have very similar predictive performance. The female screenplay writer ratio is selected in both regression models to have positive association with the Bechdel test result (coefficient estimate = 0.79) and female cast ratio (coefficient estimate = 0.042) has very high rankings in both Random Forest models (Bechdel test=6, female cast ratio=5), implying that women participation in story-writing is key to determining how women are represented on screen, outperforming the female director ratio and female producer ratio. These two variables are also selected with relatively smaller magnitude and decent ranking in the Random Forest model, due to the fact that these three variables are correlated and their effects on the response variables are adjusted for the presence of the female screenplay writer ratio. Year is also a very strong predictor of female representation, which is consistent with Fig.\ref{Fig 1} in that as time progresses women become better represented. For the female cast ratio, the number of cast members is the most influential factor (coefficient estimate=-0.19, ranking=1) as a higher number of cast members leads to a lower female cast ratio. Such an association implies that male actors are more likely to be hired than females. Many genres were identified to have better female representation, such as Romance, selected by both regression models (Bechdel test: 0.18, female cast ratio: 0.042) with decent rankings in both Random Forest models (Bechdel test: 14, female cast ratio: 6), and Horror, selected by the Bechdel test regression (coefficient estimate = 0.20) with decent rankings in both Random Forest models (Bechdel test: 11, female cast ratio: 14). Here, Horror was paid special attention due to its association with movie success in future analysis. Some genres, like Action and Crime were identified to have less female representation. Although significant correlations of budget with the female cast ratio and Bechdel test results are found in the preliminary statistical analysis, budget is not selected in either of the regression models. However, this fact can be explained by the gender preference of genres. From Fig.\ref{female_budget}, we can see that Action and Crime movies, which are identified in our models to be negatively associated with female representation, are likely to receive a much higher budget than movies that are not associated with these genres. Meanwhile, they also tend to feature more male cast members than female cast members. Romance and Horror movies (surprisingly) feature more females than males on the average. However, they also tend to receive a lower budget. The imbalance between the budgets of genres and gender preference of genres leads to the imbalance of budgets on movies with different levels of female representation. 

\begin{table*}[]
\begin{tabular}{@{}lllllll@{}}
\toprule
                                              &               & \multicolumn{2}{c}{Bechdel test} &          & \multicolumn{2}{c}{Female cast ratio}      \\ \cmidrule(lr){3-4} \cmidrule(l){6-7} 
\multicolumn{1}{c}{\multirow{2}{*}{Variable}} &               & Regression    & Random Forest    &          & Regression           & Random Forest       \\
\multicolumn{1}{c}{}                          & Test accuracy & 64.9\%        & 63.1\%           & Test MSE & $2.02\times10^{-2}$  & $1.73\times10^{-2}$ \\ \midrule
Female screenplay ratio                       &               & 0.79          & 6                &          & 0.042                & 5                   \\
Female director ratio                         &               & 0.15          & 21               &          & 1.25$\times 10^{-3}$ & 10                  \\
Female producer ratio                         &               & 0.047         & 5                &          & 0.00                 & 7                   \\
Year                                          &               & 0.67          & 2                &          & 0.016                & 4                   \\
Number of casts                               &               & 0.00          & 4                &          & -0.19                & 1                   \\
Action                                        &               & -0.26         & 8                &          & -0.018               & 8                   \\
Comedy                                        &               & 0.00          & 7                &          & 0.011                & 9                   \\
Crime                                         &               & -0.20         & 9                &          & 0.00                 & 11                  \\
Family                                        &               & 0.21          & 22               &          & 0.00                 & 34                  \\
Horror                                        &               & 0.20          & 11               &          & 0.00                 & 14                  \\
Music                                         &               & 0.04          & 28               &          & 0.00                 & 22                  \\
Romance                                       &               & 0.18          & 14               &          & 0.042                & 6                   \\
Sport                                         &               & -0.15         & 31               &          & 0.00                 & 35                  \\ \bottomrule
\end{tabular}
\caption{Regression coefficient estimates and Random Forest predictor importance ranking for the variables selected by the regularized regression models of the Bechdel test and female cast ratio. The importance of predictors in the Random Forest model is defined as the change in the Gini-impurity for classification and MSE for regression, where larger changes indicate greater importance.}
\label{Table 1}
\end{table*}

\subsection{Women Representation and Movie Success}
Setting the movie revenue/budget ratio (R/B ratio), rating, and popularity each as the response variable of an individual regression model with penalty and a Random Forest model, we include the variables of year, adult content, run time minutes, number of cast members, female cast ratio, female core crew ratio, the Bechdel test, 26 production company indicators, and 25 genre indicators as the predictors. 

The female cast ratio is selected to have strong positive association with R/B ratio (coefficient estimate=$1.19\times 10^{-3}$) and ranks as the most important predictor in the Random Forest model (ranking=1). Meanwhile, three similar genres have been identified to have high positive association with the R/B ratio: Horror (coefficient estimate=$1.69\times 10^{-3}$, ranking=7), Mystery (coefficient estimate=$2.07\times 10^{-3}$, ranking=5), and Thriller (coefficient estimate=$1.04\times 10^{-3}$, ranking=6). As we have acquired the knowledge from previous analysis that the female cast ratio is positively associated with the Horror genre, we are confident to conclude that with the presence of Horror and the other two correlated genres in the model, the female cast ratio has a positive effect on a movie's profitability after considering the effects of genres. This same set of variables is also selected for movie rating, however their associations with rating are in the opposite direction in contrast to the R/B ratio. The female cast ratio has a negative association with rating (coefficient estimate=$-6.75\times 10^{-3}$) and a very high ranking in the Random Forest model (rank=4). Horror (coefficient estimate=$-0.014$, ranking=6) and Mystery (coefficient estimate=$-0.19$, ranking=16) are identified to be negatively associated with rating. It seems that the movies featuring more women, horror, and mystery elements are likely to make more money, but receive more criticism. Neither of the female representation measures, the female cast ratio and Bechdel test result, are selected for movie popularity. However, the female cast ratio still receives a high ranking in the Random Forest model (ranking=4), possibly due to its correlation with other variables. 

\begin{figure*}
    \includegraphics[width=1.0\textwidth]{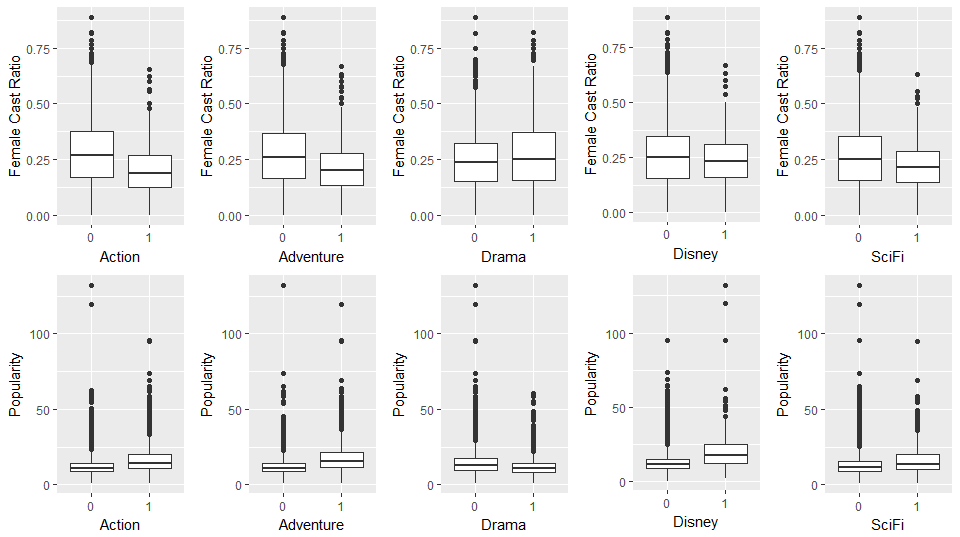}
    \caption{The female cast ratio and movie popularity distribution in the genres of Action, Adventure, Drama, Sci-Fi and production company of Disney, selected by the regression model. }
    \label{female_popularity}
\end{figure*}

Due to the fact that the revenue/budget ratio is calculated directly from budget, and that in the previous analysis we have learned that budget does not have a direct impact on female representation, we exclude budget from our model since its strong correlation with the response variable would overwhelm the effects of other variables. However, our models still demonstrate the relationship between budget and movie success. The run-time in minutes and the number of cast members are identified to have very strong association with all three aspects of movie success and receive very high ranking in the Random Forest models, and they are a proxy of a movie's budget as a higher budget leads to a larger number of cast members (Spearman correlation coefficient=0.41, p-value$<$0.0001) and a longer run time in minutes (Pearson correlation coefficient=0.26, p-value$<$0.0001). Meanwhile, our preliminary analysis shows that the female cast ratio is not significantly correlated with the R/B ratio, and our regression model identifies it to have quite a strong association possibly due to the fact that the female cast ratio is negatively associated with the budget. Movies featuring a higher female cast ratio tend to have a lower budget, which is likely to lead to a higher R/B ratio even though the revenue is not particularly high. However, such an effect is not strong enough to be detected by the correlation test. Our modeling results are consistent with the preliminary analysis for ratings in that a higher female cast ratio leads to a lower movie ratings. For movie popularity, the female cast ratio is not selected in the regression model despite the fact that the Pearson correlation test identifies them being highly correlated. This again can be explained by the confounding effect of genres and production companies. The regression model identifies some genres being positively associated with popularity, such as Action (coefficient estimate=$2.67\times 10^{-3}$, ranking=6) and Adventure (coefficient estimate=$3.19\times 10^{-3}$, ranking=5), as well as one production company Disney (coefficient estimate=$6.23\times 10^{-3}$, ranking=8). Meanwhile, it also identifies one genre being negatively associated with popularity, Drama (coefficient estimate=$-9.4\times 10^{-4}$, ranking=10). Fig.\ref{female_popularity} shows that Action, Adventure, and Sci-Fi movies, which tend to be very popular, also tend to feature fewer female cast members; Drama movies, on the other hand, feature slightly more female cast members but also tend to be less popular than other genres. Disney, known for creating strong and well-rounded female characters, in fact still features more male cast members (or male voices in animation) than female cast members in their productions. Therefore, the high popularity of Disney movies does not help associate better female representation with high movie popularity. 

\begin{table*}[]
\resizebox{\textwidth}{!}{\begin{tabular}{llllllllll}
\hline
                                              &          & \multicolumn{2}{c}{R/B ratio}              &  & \multicolumn{2}{c}{Rating}                 &  & \multicolumn{2}{c}{Popularity}            \\ \cline{3-4} \cline{6-7} \cline{9-10} 
\multicolumn{1}{c}{\multirow{2}{*}{Variable}} &          & Regression           & Random Forest       &  & Regression           & Random Forest       &  & Regression          & Random Forest       \\
\multicolumn{1}{c}{}                          & Test MSE & $1.54\times10^{-4}$  & $1.76\times10^{-4}$ &  & $6.66\times10^{-3}$  & $5.43\times10^{-3}$ &  & $1.05\times10^{-4}$ & $8.34\times10^{-5}$ \\ \hline
Female cast ratio                             &          & $1.19\times10^{-3}$  & 1                   &  & $-6.75\times10^{-3}$ & 4                   &  & 0.00                & 4                   \\
Number of casts                               &          & 0.00                 & 2                   &  & 0.014                & 3                   &  & 0.040               & 1                   \\
Run time minutes                              &          & $-5.5\times10^{-4}$  & 3                   &  & 0.29                 & 1                   &  & 0.023               & 3                   \\
Year                                          &          & 0.00                 & 4                   &  & -0.088               & 2                   &  & $2.56\times10^{-3}$ & 2                   \\
Action                                        &          & 0.00                 & 19                  &  & 0.00                 & 8                   &  & $2.67\times10^{-3}$ & 6                   \\
Adventure                                     &          & 0.00                 & 15                  &  & 0.00                 & 12                  &  & $3.19\times10^{-3}$ & 5                   \\
Animation                                     &          & 0.00                 & 27                  &  & $5.76\times10^{-3}$  & 10                  &  & 0.00                & 9                   \\
Drama                                         &          & $-5.5\times10^{-5}$  & 8                   &  & 0.015                & 5                   &  & $-9.4\times10^{-4}$ & 10                  \\
Horror                                        &          & $1.69\times10^{-3}$  & 7                   &  & -0.014               & 6                   &  & 0.00                & 24                  \\
Mystery                                       &          & $2.07\times10^{-3}$  & 5                   &  & -0.19                & 16                  &  & 0.00                & 26                  \\
SciFi                                         &          & $-1.24\times10^{-5}$ & 11                  &  & -0.018               & 15                  &  & $4.41\times10^{-4}$ & 13                  \\
Thriller                                      &          & $1.04\times10^{-3}$  & 6                   &  & 0.011                & 13                  &  & 0.00                & 17                  \\
Disney                                        &          & 0.00                 & 28                  &  & 0.00                 & 28                  &  & $6.23\times10^{-3}$ & 8                   \\ \hline
\end{tabular}}
\caption{Regression coefficient estimates and Random Forest predictors importance ranking for variables selected by the regularized regression models of movie revenue/budget ratio, rating and popularity.}
\label{table 2}
\end{table*}

\section{Conclusions}
Overall, our findings regarding female representation indicate that female representation is critically influenced by the work of female crew members, especially female screenplay writers, in the film-making process. In addition, it is also evolving throughout time as a result of other factors outside movies and thus not included in the dataset (i.e. the variable year). Our findings also reflect some difficulties actresses face, for example, male actors are more likely to be hired than females, genres that tend to receive higher budgets also prefer male actors to portray the stories, and so on. From our investigation into the relationship between female representation and movies' success, we discover that compared to the Bechdel test result, the proposed female cast ratio is more directly linked to a movie's success as it is often selected multiple times. In comparison, the Bechdel test result is not selected once. Moreover, movies featuring more women tend to have a lower budget which leads to a higher revenue/budget return. However, they also suffer from more criticism, possibly due to the low budget invested. Considering that the number of cast members and run-time minutes are both selected to have a very high positive impact on a movie's rating and that they are both a proxy of the budget, this is a very plausible explanation. The female cast ratio is not directly linked to a movie's popularity. However, we have also discovered that genres likely to be popular, such as adventure and action, also tend to feature fewer female cast members. Disney has made contributions to better female representation on the big screen and its productions are often very popular. However, their productions still feature more males than females. 

Our findings have demonstrated that the difficulties women in the film industry face are both upstream and downstream: female filmmakers, especially female screenplay writers, are instrumental for movies to have better female representation, but the percentages of female filmmakers are very low; only 6.0\% of directors, 9.7\% of producers, and 12.2\% of screenplay writers are female. Meanwhile, lower budgets are provided to support the movies that could tell good stories about women, thus causing the films to receive more criticism. The demand of better female representation from viewers is also not strong enough to press the film industry for change, as movies that have poor female representation can still be very profitable. 

\section{Discussions}

Our study provides a larger picture of female representation in movies and how it is perceived by the audience over time. Unfortunately, underrepresentation of women in movies is not the only difficulty women are facing. Apart from the low ratio of female cast members, portrayal of women in stereotypical ways that reflect and sustain socially endorsed views of genders, and depictions of relationships between men and women that emphasize traditional roles and normalize violence against women, are another two important themes of how media portray gender ~\cite{wood1994gendered}. Although our newly-defined measure of female representation, the female cast ratio, has shown to be more successful in evaluating movie success than the Bechdel test, it also has limitations that fail to further explain how women are portrayed in movies. For example, our findings indicate that horror movies like to feature more women than men, and the reason behind this phenomenon is that the audience enjoys the victimization of women more than the victimization of men ~\cite{horrormovie}. In other words, the root of "favoritism" toward women in horror movies is still that people want to see women being helpless and passive rather than men. Another genre identified by our analysis that favors women, Romance, is also likely to portray women in a stereotypical way. A linguistics study of the movie "The Best of Me" shows some differences in the expressions between men and women, in that men prefer to use commanding directives while women use directives for suggesting or requesting, and that men tend to use swear words to express anger and women tend to use swear words to express bad feelings ~\cite{bestofme}. Even in a movie that targeted the female audience, women were portrayed as docile and submissive compared to men. Meanwhile, some studies have used dialogue speaking time to evaluate how much women talk in movies and discovered that even in some female-led movies (such as Disney princess movies), the lead female's speaking time could be outnumbered by male cast members ~\cite{FilmDialogue}. Whether they speak to express their ideas, feelings, or make commands, men speak more than women and their influence on the audience is ultimately stronger. 

We will also consider many other factors that can influence the image of a character in movies in our future studies to help us better understand how gender works in shaping a character's image and how they influence the audience's perception of movies. For example, race and ethnicity are both important factors that impact the audience's expectation for a character. For example, studies have shown that African American music videos were significantly more likely to portray sexual content and sexualized female characters than White videos ~\cite{turner2011sex}. In addition, the first African American performer to win the Academy Award was a female, and she won much earlier than her male counterparts. The reason behind the phenomenon that minority women are more popular than minority men is possibly that women and minorities were traditionally both considered to be submissive to white males, thus a minority woman portrayed in such a way is more acceptable to the mainstream than a minority man. Besides race and ethnicity, age is another important factor. While in movies both women and men in their 60s and older are dramatically underrepresented compared to their representation in the U.S. population ~\cite{lauzen2005maintaining}, older actresses experience greater difficulty in finding a job than older actors even for celebrities~\cite{treme2013celebrity}. The majority of male characters are in their 30s and 40s, and the majority of female characters are in their 20s and 30s. For male characters, leadership and occupational power increase with age; however, as female characters age, they are less likely to have goals ~\cite{lauzen2005maintaining}. This observation coincides with the mainstream expectation of genders in that men hold authority and leadership which usually increases with age, while women are in more subordinate roles and often sexualized, which is why younger women are preferred. 

In conclusion, we hope our findings send a message to the film-making industry that movies with better female representation are more likely to succeed. Improving female representation in movies requires both upstream effort and downstream effort: female crew and cast members contribute to better female representation in movies, and positive feedback from the audience encourages investment into movies to hire more female crew and cast members. Therefore, we encourage the film-making industry to employ more women as a start. Men have been the leading voices of story-telling in movies for decades, so we believe that stories from women's perspectives are worth exploring and would attract larger audiences. 

\section*{Acknowledgement}
The authors would like to thank Joyce Luo and William Artman for their help with writing.

\bibliographystyle{aaai}
\bibliography{bib}

\begin{thebibliography}{}

\bibitem[\protect\citeauthoryear{Anderson and Daniels}{2016}]{FilmDialogue}
Anderson, H., and Daniels, M.
\newblock 2016.
\newblock Film dialogue from 2,000 screenplays, broken down by gender and age.
\newblock \url{https://pudding.cool/2017/03/film-dialogue/index.html}.

\bibitem[\protect\citeauthoryear{Berkman, Garland, and
  VanSteinberg}{2017}]{Duke}
Berkman, S.; Garland, S.; and VanSteinberg, A.
\newblock 2017.
\newblock Quantified feminism and the {B}echdel test.
\newblock Technical report, Duke University.

\bibitem[\protect\citeauthoryear{Collins}{2011}]{Collins2011}
Collins, R.~L.
\newblock 2011.
\newblock Content analysis of gender roles in media: Where are we now and where
  should we go?
\newblock {\em Sex Roles} 64(3):290--298.

\bibitem[\protect\citeauthoryear{Haraldsson and
  W{\"a}ngnerud}{2019}]{womenpolitical}
Haraldsson, A., and W{\"a}ngnerud, L.
\newblock 2019.
\newblock The effect of media sexism on women’s political ambition: evidence
  from a worldwide study.
\newblock {\em Feminist Media Studies} 19(4):525--541.

\bibitem[\protect\citeauthoryear{Heldman, Frankel, and Holmes}{}]{heldman2016}
Heldman, C.; Frankel, L.~L.; and Holmes, J.
\newblock “hot, black leather, whip” the (de) evolution of female
  protagonists in action cinema, 1960--2014.
\newblock {\em Sexualization, Media, \& Society} 2(2).

\bibitem[\protect\citeauthoryear{Ho}{1995}]{ho1995random}
Ho, T.~K.
\newblock 1995.
\newblock Random decision forests.
\newblock In {\em Proceedings of 3rd international conference on document
  analysis and recognition}, volume~1,  278--282.
\newblock IEEE.

\bibitem[\protect\citeauthoryear{Lauzen and
  Dozier}{2005}]{lauzen2005maintaining}
Lauzen, M.~M., and Dozier, D.~M.
\newblock 2005.
\newblock Maintaining the double standard: Portrayals of age and gender in
  popular films.
\newblock {\em Sex roles} 52(7-8):437--446.

\bibitem[\protect\citeauthoryear{Lauzen}{2012}]{lauzen2012celluloid}
Lauzen, M.
\newblock 2012.
\newblock The celluloid ceiling: Behind the scenes employment of women on the
  top 250 films of 2013.
\newblock {\em Women’s Media Center} 15:2012.

\bibitem[\protect\citeauthoryear{Lindner and Schulting}{}]{lindner2017movies}
Lindner, A.~M., and Schulting, Z.
\newblock How movies with a female presence fare with critics.
\newblock {\em Socius} 3.

\bibitem[\protect\citeauthoryear{Lindner, Lindquist, and
  Arnold}{2015}]{lindner2015million}
Lindner, A.~M.; Lindquist, M.; and Arnold, J.
\newblock 2015.
\newblock Million dollar maybe? {T}he effect of female presence in movies on
  box office returns.
\newblock {\em Sociological Inquiry} 85(3):407--428.

\bibitem[\protect\citeauthoryear{Micic}{2015}]{micic2015female}
Micic, Z.
\newblock 2015.
\newblock Female interactions on film-beyond the {B}echdel test: A quantitative
  content analysis of same-sex-interactions of top 20 box office films.
\newblock
  \url{http://www.divaportal.org/smash/get/diva2:821217/FULLTEXT01.pdf}.

\bibitem[\protect\citeauthoryear{Murphy}{2015}]{murphy2015role}
Murphy, J.~N.
\newblock 2015.
\newblock {\em The role of women in film: Supporting the men--An analysis of
  how culture influences the changing discourse on gender representations in
  film}.
\newblock Undergraduate {H}onor {T}hesis, Department of Journalism, University
  of Arkansas.

\bibitem[\protect\citeauthoryear{Sulastri, Laila, and Hum}{2019}]{bestofme}
Sulastri, S.; Laila, M.; and Hum, M.
\newblock 2019.
\newblock {\em Characterizing Men And Women Language In {T}he {B}est Of {M}e
  Movie}.
\newblock Ph.D. Dissertation, Universitas Muhammadiyah Surakarta.

\bibitem[\protect\citeauthoryear{Tamborini, Stiff, and
  Zillman}{1987}]{horrormovie}
Tamborini, R.; Stiff, J.; and Zillman, D.
\newblock 1987.
\newblock Preference for graphic horror featuring male versus female
  victimization.
\newblock {\em Human Communication Research} 13(4):529--552.

\bibitem[\protect\citeauthoryear{Thomas}{2017}]{201610}
Thomas, A.
\newblock 2017.
\newblock Women only said 27\% of the words in 2016’s biggest movies.
\newblock \url{https://www.freecodecamp.org/news/}.

\bibitem[\protect\citeauthoryear{Tibshirani}{1996}]{tibshirani1996regression}
Tibshirani, R.
\newblock 1996.
\newblock Regression shrinkage and selection via the lasso.
\newblock {\em Journal of the Royal Statistical Society: Series B
  (Methodological)} 58(1):267--288.

\bibitem[\protect\citeauthoryear{Treme and Craig}{2013}]{treme2013celebrity}
Treme, J., and Craig, L.~A.
\newblock 2013.
\newblock Celebrity star power: Do age and gender effects influence box office
  performance?
\newblock {\em Applied Economics Letters} 20(5):440--445.

\bibitem[\protect\citeauthoryear{Turner}{2011}]{turner2011sex}
Turner, J.~S.
\newblock 2011.
\newblock Sex and the spectacle of music videos: An examination of the
  portrayal of race and sexuality in music videos.
\newblock {\em Sex Roles} 64(3-4):173--191.

\bibitem[\protect\citeauthoryear{Wood}{1994}]{wood1994gendered}
Wood, J.~T.
\newblock 1994.
\newblock Gendered media: The influence of media on views of gender.
\newblock {\em Gendered lives: Communication, gender, and culture} 9:231--244.

\end{thebibliography}
\end{document}